# A Faster Randomized Algorithm for Vertex Cover: An Automated Approach

## Katie Clinch
School of Mathematics and Physics, University of Queensland, Brisbane, Australia

## Serge Gaspers
School of Computer Science and Engineering, UNSW Sydney, Australia

## Tao Zixu He
Manning College of Information and Computer Sciences, University of Massachusetts Amherst, United States of America

## Simon Mackenzie
School of Computer Science and Engineering, UNSW Sydney, Australia

## Tiankuang Zhang
School of Computer Science and Engineering, UNSW Sydney, Australia

## Abstract

This work introduces two techniques for the design and analysis of branching algorithms, illustrated through the case study of the VERTEX COVER problem. First, we present a method for automatically generating branching rules through a systematic case analysis of local structures. Second, we develop a new technique for analyzing randomized branching algorithms using the Measure & Conquer method, offering greater flexibility in formulating branching rules. By combining these innovations with additional techniques, we obtain the fastest known randomized algorithms in different parameters for the VERTEX COVER problem on graphs with bounded degree (up to 6) and on general graphs. For example, our algorithm solves VERTEX COVER on subcubic graphs in $O^*(1.07625^n)$ time and $O^*(1.13132^k)$ time, respectively. For graphs with maximum degree 4, we achieve running times of $O^*(1.13735^n)$ and $O^*(1.21103^k)$, while for general graphs we achieve $O^*(1.25281^k)$.

**2012 ACM Subject Classification** Theory of computation, Design and analysis of algorithms

**Funding** *Katie Clinch*: Australian Research Council (project DP210103849)
*Serge Gaspers*: Australian Research Council (project DP210103849)
*Simon Mackenzie*: Australian Research Council (project DP210103849)
*Tiankuang Zhang*: Australian Research Council (project DP210103849)

**Acknowledgements** We thank the authors of [12, 24] for their assistance in reconstructing their running time analyses. We thank Daniel Lokshtanov for preliminary discussions of randomized branching.

## 1 Introduction

Exact branching algorithms are among the most effective tools for tackling NP-hard problems [20], yet they are still largely *hand crafted*: each new problem requires bespoke case analyses and carefully tuned measures [8]. This artisanal process limits portability and slows progress. We argue that much of this work can be automated. We present a framework that (i) systematically explores local configurations, (ii) streamlines the search by collapsing equivalences, and (iii) generates high quality deterministic or randomized branching rules by solving `LP/ILP` formulations that directly optimize measure progress.

We demonstrate the approach on VERTEX COVER, improving state-of-the-art bounds on both bounded degree and general graphs. In particular, for general graphs parameterized by solution size $k$, we reduce the best known running time to $O^*(1.25281^k)$, with further improvements under degree caps (see Table 1). Methodologically, a central ingredient is a randomized variant of Measure & Conquer (M&C) that enables tight analyses of probabilistic branching, filling a gap in the literature where, to the best of our knowledge, M&C has been used exclusively for deterministic algorithms.

**The Vertex Cover problem**

The VERTEX COVER problem has been studied for over fifty years [9]. Given a graph $G = (V, E)$ and $k \in \mathbb{Z}_{\geq 0}$, the task is to decide whether there exists $C \subseteq V$ of size at most $k$ covering all edges. Two main algorithmic paradigms dominate exact algorithms: exponential time algorithms and fixed parameter tractable (FPT) algorithms. The fastest exponential time algorithm runs in $O^*(1.19951^n)$ time [23], while VERTEX COVER is FPT with respect to parameters such as $k$ [4], treewidth (Folklore; e.g., [3]), and $k - \texttt{LP}(G)$ [15]. Prior to our work, the fastest FPT algorithm for general graphs parameterized by $k$ had a running time of $O^*(1.25284^k)$ [12]. Beyond general graphs, structure helps: VERTEX COVER is polynomial time solvable on bipartite graphs [19], and bounded degree instances often admit faster branching algorithms even though the problem remains NP-complete already for subcubic graphs.

**Automating branching design**

The idea of *automatic algorithm generation* has gained traction in the last two decades. Gramm et al. [11] pioneered the approach for CLUSTER EDITING, later extended by Tsur [18]; Fedin and Kulikov [5] proposed a generator for SAT; and more recently, Červenỳ and Suchỳ [2] adapted the paradigm to $d$-PATH VERTEX COVER, obtaining a $O^*(1.3294^k)$ algorithm. In parallel, randomization has repeatedly yielded faster exact algorithms [13, 17], with Monotone Local Search improving running times for dozens of NP-complete problems [6]. Our work unifies these perspectives: we automatically *generate* randomized branching algorithms and *analyze* them via a generalized M&C framework.

**Key contributions**

1. **Randomized Measure & Conquer.** We extend Measure & Conquer to analyze the running time of randomized algorithms (Lemma 2). This generalization broadens the applicability of M&C, allowing it to handle probabilistic branching rules effectively.
2. **Automated rule generation via `LP/ILP`.** We introduce a novel framework, cast rule selection and weighting as an optimization problem that directly maximizes measure decrease, naturally supporting both deterministic and randomized rules.

3. **Improved running times for Vertex Cover.** Our generated algorithms sharpen the best known parameterized bounds on general graphs to $O^*(1.25281^k)$ and improve several bounded degree cases; see Table 1 for a summary.

| Max. Degree | Parameter | New Bound | Previous Bound |
|---|---|---|---|
| $\Delta \leq 3$ | $n$ | $O^*(1.07625^n)$ | $O^*(1.08351^n)$ |
|  | $k$ | $O^*(1.13132^k)$ | $O^*(1.14416^k)$ |
| $\Delta \leq 4$ | $n$ | $O^*(1.13735^n)$ | $O^*(1.13760^n)$ |
|  | $k$ | $O^*(1.21103^k)$ | $O^*(1.21131^k)$ |
| $\Delta \leq 5$ | $n$ | — | $O^*(1.17354^n)$ |
|  | $k$ | $O^*(1.24382^k)$ | $O^*(1.24394^k)$ |
| $\Delta \leq 6$ | $n$ | — | $O^*(1.18922^n)$ |
|  | $k$ | $O^*(1.25210^k)$ | $O^*(1.25214^k)$ |
| $\Delta \leq 7$ | $n$ | — | $O^*(1.19698^n)$ |
| $\Delta \leq n-1$ | $n$ | — | $O^*(1.19951^n)$ |
|  | $k$ | $O^*(1.25281^k)$ | $O^*(1.25284^k)$ |

**Table 1** Known and new results for Vertex Cover. New bounds are proved in this work; previous bounds from [21, 24, 22, 23, 12].

*Remarks.* Beyond these results, our framework substantially reduces the human effort in branching analyses: tasks left to the analyst are minimal (instance partitioning is straightforward and simplification rules are classical), enabling studies that were previously computationally prohibitive; incorporating richer hand crafted insights could tighten the bounds even further. Moreover, the randomized Measure & Conquer method and the automated generation framework are modular and can be applied independently to other problems with minimal adaptation. By casting rule selection and weighting as LP/ILP formulations, the framework systematically searches a large design space and targets high efficiency branching rules. Finally, the instance partitioning mechanism offers a clean hook for non-local structure, special substructures can be handled manually and their consequences encoded as assertions in the subspace definitions.

**Comparison to prior work**

Our framework diverges from previous automated approaches along three key axes. *(i) Enumeration.* Gramm et al. and Červenỳ–Suchỳ grow local configurations by adding *vertices*, while Tsur allows partial information and adds at most two edges and one vertex at a time [2, 11, 18]. In contrast, we expand configurations by adding one *edge* at a time. Together with support for partially specified edges and an explicit maximum degree bound, this edge-centric growth sharply reduces the number of candidate configurations to explore. *(ii) Redundancy control.* Fedin–Kulikov and Tsur [5, 18] do not deduplicate; Gramm et al. and Červenỳ–Suchỳ [11, 2] apply full graph isomorphism, with the latter also using subgraph isomorphism, potentially leading to substantial computational cost. We instead partition the instance space into structured subspaces and merge isomorphic local configurations, eliminating duplicates while avoiding the overhead of frequent subgraph isomorphism queries. *(iii) Rule generation.* Gramm et al. exhaustively enumerate branches and pick the best; Fedin–Kulikov select from a fixed catalogue; and Červenỳ–Suchỳ and Tsur generate correctness guaranteed rules and simplify them post hoc [11, 5, 2, 18]. We instead formulate rule selection and weighting as a single `LP/ILP` optimization problem that directly maximizes measure progress and seamlessly supports randomized branching. This allows us to consider a much larger space of branching rules, especially for randomized branching where it suffices to solve linear programs.

**Roadmap**

The structure of this paper is organized as follows. Section 2 introduces the key notations and foundational concepts. In Section 3, we present an approach for applying M&C to randomized algorithms. Section 4 details the automated algorithm generation framework and demonstrates its effectiveness by generating randomized algorithms for VERTEX COVER. In Section 5, we evaluate the performance of the generated algorithms and apply existing techniques to establish new exact and parameterized running time upper bounds for VERTEX COVER on both general and bounded degree graphs. Finally, Section 6 summarizes our contributions and outlines promising directions for future research. Unless otherwise specified, omitted proofs can be found in Appendix B.

## 2 Preliminaries

We review some key concepts about graphs, branching algorithms, and the M&C method.

**Graph theory**

Let $G = (V, E)$ be an undirected graph where $n = |V|$. The *degree* of a vertex $v \in V$ is denoted $\deg_G(v)$. The *maximum degree* of $G$, denoted by $\Delta(G)$, is given by $\Delta(G) = \max\{\deg_G(v) \mid v \in V\}$. An *induced subgraph* $G[V'] = (V', E')$ of $G$ consists of the vertex set $V' \subseteq V$ and edge set $E' = \{\{u, v\} \in E \mid u, v \in V'\}$. For a subset $V' \subseteq V$, the induced graph $G[V \setminus V']$ is also denoted by $G - V'$. For a single vertex $v \in V$, we write $G - \{v\}$ simply as $G - v$. For a vertex $v \in V$, the *open neighborhood* of $v$ is denoted by $N(v) = \{u \in V \mid \{u, v\} \in E\}$ and the *closed neighborhood* is $N[v] = N(v) \cup \{v\}$. The size of the minimum vertex cover of $G$ is denoted by $\mathtt{VC}(G)$.

**Simple branching algorithms**

In this paper, we restrict our attention to a constrained family of branching algorithms called *simple branching algorithms*. A simple branching algorithm is a recursive algorithm that explores the search space of possible solutions (certificates) by iteratively refining the space. It does not rely on any information from previously solved instances while solving the current one. To construct such an algorithm, *branching rules* are applied, which select a local structure within the instance, generate smaller subproblems, and make decisions about parts of the solution, such as whether a specific vertex should be included in the solution for VERTEX COVER. The algorithm then recursively explores these sub-instances, looking for valid certificates. As it progresses, it builds a *search tree*, where each node represents a choice, and the time spent at each recursive step is typically polynomial. The overall running time is bounded by the size of the search tree.

**Measure & Conquer**

The key tool in the M&C method [7] is the concept of a *measure*, which is central to both generating the branching rules, and obtaining a corresponding bound on the running time. Lemma 1 gives the constraints that a measure must satisfy in order to be used in the analysis of a branching algorithm.

▶ **Lemma 1** (Lemma 2.5 in [10]). *Let $\mathcal{A}$ be a branching algorithm for a decision problem $\Pi$, and let $\mu(\cdot)$ be a non-negative measure for instances of $\Pi$. For any instance $I$ of $\Pi$, $\mathcal{A}$ either solves $I$ in constant time if $\mu(I) = 0$, or reduces $I$ to sub-instances $I_1, \ldots, I_k$ such that,*

$$\sum_{i=1}^{k} 2^{\mu(I_i)} \leq 2^{\mu(I)}. \tag{1}$$

*If $\mathcal{A}$ recursively solves all sub-instances in polynomial depth and combines their solutions to solve $I$, using only polynomial time for the reduction and backtracking (excluding the recursive calls), then $\mathcal{A}$ solves $I$ in $O^*(2^{\mu(I)})$ time.*

In M&C, we use the measure to track certain characteristics of an instance to monitor the progress made by a branching algorithm. By doing so, we can balance slower branching rules with faster ones to optimize the worst case running time. In this work, we use the soft-$O$ notation $O^*$, which is similar to $O$ but suppresses polynomial factors in the input size in asymptotic analysis.

## 3    Measure & Conquer for Randomized Algorithms

In this section, we present an approach for applying M&C to randomized algorithms. Specifically, we consider a randomized variant of the simple branching algorithm, where the algorithm randomly selects one sub-instance to solve recursively according to a predefined probability distribution, and then solves the original instance based on the solution of the chosen sub-instance. For any no-instance, the algorithm consistently returns No, whereas for any yes-instance, it returns Yes with a positive success probability. We refer to this modified procedure as a *randomized branching algorithm*.

A randomized branching algorithm can offer greater flexibility than its deterministic counterpart, especially when there are multiple leaves in the search tree that provide solutions. For instance, consider the Vertex Cover problem: if a yes-instance contains a 4-clique, then at least three vertices of this clique must be included in the vertex cover. A typical 2-way deterministic branching rule has one branch that includes the first vertex and another branch that discards the first and includes the other 3 vertices into the vertex cover. This yields a running time of $f(k) = f(k-1) + f(k-3)$, which solves to $f(k) \in O^*(1.46558^k)$. Another option is to branch on each vertex of the 4-clique, generating four sub-instances with three vertices added into the vertex cover in each. This yields a running time of $f(k) = 4f(k-3)$, which solves to $f(k) \in O^*(1.58741^k)$. By contrast, a randomized branching algorithm can select one vertex uniformly at random. Since the expected number of trials to choose a correct vertex is $\frac{4}{3}$, the resulting running time is $f(k) = \frac{4}{3}f(k-1)$, which solves to $f(k) \in O^*(1.33334^k)$.

Moreover, this idea extends naturally to more general settings. To rigorously analyze the expected running time of such randomized branching algorithms, we present the following lemma, which formalizes how M&C can be used to design and analyze these algorithms.

▶ **Lemma 2.** *Let $\mathcal{A}_r$ be a randomized branching algorithm for a decision problem $\Pi$, and let $\mu(\cdot)$ be a non-negative measure for instances of $\Pi$. For any instance $I$ of $\Pi$, $\mathcal{A}_r$ either solves $I$ in constant time if $\mu(I) = 0$, or reduces $I$ to sub-instances $I_1, \ldots, I_k$ with corresponding weights $w_1, \ldots, w_k$. If $I$ is a yes-instance, the sum of the weights of the yes-sub-instances is at least one; otherwise, all sub-instances are no-instances. A sub-instance $I_i$ is selected at random with probability at least $w_i \cdot 2^{\mu(I_i) - \mu(I)}$, subject to the constraint*

$$\sum_{i=1}^{k} w_i \cdot 2^{\mu(I_i)} \leq 2^{\mu(I)}. \tag{2}$$

*If $\mathcal{A}_r$ recursively solves the chosen sub-instance $I_i$ in polynomial depth and uses its solution to solve $I$, with polynomial time required for the reduction and backtracking, then $\mathcal{A}_r$ runs in polynomial time on every instance. Moreover, for any yes-instance $I$, $\Pr[\mathcal{A}_r \text{ solves } I] \geq 2^{-\mu(I)}$.*

The randomized branching algorithm is presented in Algorithm 1, where the function `Branch` reduces an instance to sub-instances with corresponding weights that satisfy Equation 2. The algorithm $\mathcal{A}_r$ has a success probability of at least $2^{-\mu(I)}$. In order to achieve a constant success probability, we run the algorithm $2^{\mu(I)}$ times.

**Algorithm 1** `RSearch`$(I, \mu)$

**Input:** An instance $I$, and a measure $\mu(\cdot)$ for instances of $\Pi$
**Output:** YES with probability at least $2^{-\mu(I)}$ if $I$ is a YES-instance, NO otherwise

1  **if** $\mu(I) = 0$ **then**
2  |  **return** *Solve(I)*; // Solve the instance in constant time
3  $(I_1, w_1), \ldots, (I_k, w_k) \leftarrow \texttt{Branch}(I)$;
4  $p^* \leftarrow \sum_{i=1}^{k} w_i 2^{\mu(I_i)}$;
5  **foreach** $I_i \in I_1, \ldots, I_k$ **do**
6  |  $p_i \leftarrow w_i 2^{\mu(I_i)} / p^*$ ; // ensure the probability sum to 1
7  Randomly choose $i \in \{1, \ldots, k\}$, with probability $p_i$;
8  **return** *RSearch*$(I_i, \mu)$;

Our Randomized Measure & Conquer fills an interesting gap in the literature. The straightforward way to generalize a Measure & Conquer analysis of a branching algorithm to a randomized branching algorithm sets all weights to 1; this then gives the same running time for the randomized and deterministic variants. Except for simple cases, discovering and reasoning about weighted randomized branching rules seems unintuitive; a systematic exploration of weighted branching rules and optimization (via LPs) appears necessary to design randomized branching algorithms that benefit meaningfully from the additional flexibility of the weights of the branches.

The randomness introduces additional flexibility but comes at the cost of restricting the computation to simple branching algorithms, where the computation may not depend on the result of previous branches. When Monotone Local Search is combined with a M&C analysis, we can think of adding a randomized branching rule to the M&C algorithm: if the solution size is above some threshold, add a random vertex/element to the solution. Integrating the analysis of this particular randomized branching rule into Measure & Conquer has been explored in [14]. However, even in this context where the random component of the algorithm emerges naturally, it can generally be derandomized. Designing randomized branching algorithms that beat deterministic ones both in simplicity and in running time is a wide open and challenging area of research. However, what we show in this paper is that the automatic search for algorithms benefits from the inclusion of randomness, as this allows us to relax the ILP component of the method to an LP. This LP relaxation not only makes the optimization more tractable computationally but also allows for exploring a richer space of simple branching strategies. Casting the automation in that richer space provides the dual benefits of facilitating the finding of faster algorithms but also deepening our understanding of how randomness may be exploited in branching algorithms.

## 4 Automated Framework

This section presents a framework for automatically generating branching algorithms (including randomized ones), illustrated by an application to Vertex Cover on subcubic graphs.

A branching algorithm can be conceptualized as one that maintains a set of branching rules, each associated to a specific property of input instances. When the algorithm encounters an instance, it identifies the property that matches the instance and applies the corresponding rule. The algorithm solves a problem successfully if, for every instance, a matching property exists and the associated rule functions correctly for all instances with that property. In our framework, these properties are captured by local configurations, which decompose instances into smaller and manageable components for fine-grained analysis.

▶ **Definition 3** (Local configuration). *We define a local configuration for the Vertex Cover problem as a tuple $L = (H, D)$, where $H = (V, E)$ is a graph and $D$ is a function that maps each vertex to the number of incomplete edges incident to it. Intuitively, this describes a local structure in which a subgraph $H$ exists within a larger graph, and for every vertex $v \in V$, exactly $D(v)$ edges incident to $v$ in the full graph are not included in $H$[1]. The set of vertices with incomplete edges, known as the* boundary set, *is denoted by $\delta(L) = \{v \in V \mid D(v) \neq 0\}$.*

▶ **Definition 4** (Expansion). *Let $L = (H, D)$ and $L' = (H', D')$ be two local configurations with $H = (V, E)$ and $H' = (V', E')$. We say that $L'$ is an* expansion *of $L$ (denoted $L' \geq L$) if there exists an injective mapping $\phi : V \to V'$ such that, for all $u, v \in V$, $\{u, v\} \in E$ implies $(\phi(u), \phi(v)) \in E'$. Moreover, for each $v \in V$, the incomplete edge function must satisfy $D'(\phi(v)) - D(v) = \deg_{H'}(\phi(v)) - \deg_H(v) \geq 0$. In this way, a local configuration captures a microstructure that is preserved across all its expansions, thereby serving as a fundamental building block for analyzing larger instances. For simplicity, and without loss of generality, we assume throughout that $\phi$ is the identity mapping, allowing us to treat $V$ as elements of $V'$ in our analysis.*

Note that a graph $G = (V, E)$ can be viewed as a local configuration $L = (G, D)$ with $D(v) = 0$ for all $v \in V$.

Besides the notion of local configuration, we also introduce the following simplification rules for an instance consisting of a graph $G = (V, E)$ and the vertex cover size $k$.

1. **Isolated Vertex Rule:** If a vertex $v$ has degree 0, remove $v$ from $G$.
2. **Degree-1 Vertex Rule:** If a vertex $v$ has degree 1, remove $N[v]$ from $G$ and decrement $k$ by 1.

---

[1] These edges are incident to $v$ in the full graph and connect $v$ either to vertices inside $V$ (but excluded from $E$) or to vertices outside $V$.

3. **Degree-2 Triangle Rule:** If a vertex $v$ has degree 2 and its neighbors $u$ and $w$ are adjacent, remove $N[v]$ from $G$ and decrement $k$ by 2.

4. **Degree-2 Chain Rule:** If two adjacent vertices $u$ and $v$ both have degree 2, remove $u$ and $v$ from $G$, add an edge between their remaining neighbors, and decrement $k$ by 1.

5. **Alternating Cycle Rule:** If there exists an even-length simple cycle $c$ of length $l$ with alternating degree-2 and higher-degree vertices, remove all vertices in $c$ from $G$ and decrement $k$ by $\frac{l}{2}$.

If a simplification rule can be applied to the instance, we use it to obtain a simplified sub-instance. If no rule is applicable, we attempt to find a local configuration that can be expanded to the instance. This process assumes that none of the structures targeted by the simplification rules remain in the instance.

This section is organized as follows. In Subsection 4.1, we introduce the concept of an *expansion cover*, which reduces the task of finding efficient branching rules from a local configuration to larger, more comprehensive ones. Next, Subsection 4.2 describes the measure used for solving VERTEX COVER, details its associated constraints, and presents a formula to estimate how the worst-case measure changes when a branch is applied. Subsequently, Subsection 4.3 focuses on generating efficient branching rules for local configurations while also discussing various optimization techniques. Finally, Subsection 4.4 offers a comprehensive overview of the framework, along with its pseudocode.

## 4.1 Expansion cover

Since a local configuration captures the common substructure of all its expansions, and every instance can be represented as a local configuration without boundary vertices, a local configuration effectively represents an instance space. For example, a local configuration consisting of a single vertex and 3 incomplete edges represents all graphs containing a vertex of degree-3. A *cover* of a local configuration $L$ is defined as a set of local configurations $\mathcal{L}$ such that the instance space of $L$ is a subset of the union of the instance spaces of all elements in $\mathcal{L}$.

By finding a cover of a local configuration, the task of determining an efficient branching rule for that configuration reduces to finding efficient branching rules for its covering configurations. Given a local configuration $L = (H, D)$ with a boundary vertex $v \in \delta(L)$, we can construct a cover by examining the incident vertex $u$ of one of $v$'s incomplete edges. Specifically, we distinguish two cases for the endpoint $u$ of this incomplete edge: either $u$ lies in $V$ or $u$ lies outside $V$. We then consider all possible local configurations arising in each case.

Algorithm 2 outlines an algorithm for expanding a local configuration of graphs of maximum degree $\Delta$. The algorithm begins by selecting a boundary vertex $v$ with the minimum degree and the fewest incomplete edges. For each $u_i$ in $\delta(L) \setminus \{v\}$, the algorithm creates a new local configuration by instantiating one of $v$'s incomplete edges to connect $v - u_i$. Additionally, it generates $\Delta$ more configurations by adding a new vertex $u$ (of degree $i$) to $V$ for each $i \in \{1, \ldots, \Delta\}$, and connecting $u$ to $v$. The union of these newly generated local configurations forms a cover of $L$. Finally, if a local configuration of constant size has no boundary vertices (i.e., $\delta(L) = \emptyset$), it is an isolated connected component that can be solved in constant time.

### ▰ Algorithm 2 Expand$(L, \Delta)$

**Input:** A local configuration $L = (H, D)$, $H = (V, E)$ and an integer $\Delta$ denoting the maximum degree
**Output:** A set of local configurations $\mathcal{L}$

1 $\mathcal{L} \leftarrow \{\emptyset\}$;
2 let $v \in \{v \in \delta(L) \mid \forall_{u \in \delta(L)} D(v) < D(u) \vee (D(v) = D(u) \wedge \deg(v) \leq \deg(u))\}$;
3 **foreach** $u \in \delta(L)$ **do**
4 $\quad D_u(x) = \begin{cases} D(x) - 1 & \text{if } x \in \{u, v\} \\ D(x) & \text{otherwise} \end{cases}$;
5 $\quad H_u \leftarrow (V, E \cup \{\{u, v\}\})$;
6 $\quad L_u \leftarrow (H_u, D_u)$;
7 $\quad \mathcal{L} \leftarrow \mathcal{L} \cup \{L_u\}$;
8 **foreach** $d \in \{1, \ldots, \Delta\}$ **do**
9 $\quad u$ is a new vertex;
10 $\quad D_u(x) = \begin{cases} d - 1 & \text{if } x = u \\ D(x) - 1 & \text{if } x = v \\ D(x) & \text{otherwise} \end{cases}$;
11 $\quad H_u \leftarrow (V \cup \{u\}, E \cup \{\{u, v\}\})$;
12 $\quad L_u \leftarrow (H_u, D_u)$;
13 $\quad \mathcal{L} \leftarrow \mathcal{L} \cup \{L_u\}$;
14 **return** $\mathcal{L}$;

## 4.2 Measure

We combine M&C with branch cost (discussed in Subsubsection 4.3.2) to enable the framework to better track the progress of branching rules, resulting in fewer layers of expansion before termination. To apply M&C, we first define the measure as:

$$\mu = \alpha k + \beta_1 n_1 + \beta_2 n_2 + \beta_3 n_3, \tag{3}$$

where $k$ is the size of the vertex cover, $n_i$ denotes the number of vertices of degree-$i$ for $i \in \{1, 2, 3\}$, and $\alpha, \beta_1, \beta_2, \beta_3$ are the weights assigned to these components.

**Constraints**

The weights $\alpha, \beta_1, \beta_2, \beta_3$ are manually determined and form part of the framework's input. By Lemma 1, we must choose weights so that no simplification rule increases the measure. To enforce this, we identify a set of basic *atomic* operations, each of which places one vertex into the cover (reducing $k$ by 1). We then express each simplification rule as a sequence of these atomic operations (all except Rules 1 and 4 can be so expressed). Rule 1 is benign (it leaves the measure unchanged), and Rule 4 is handled separately, with a special-case analysis of its measure change. For each atomic operation, we construct constraints to ensure that the measure change is non-positive. A detailed list of operations and the corresponding constraints can be found in Appendix A.

To generate algorithms parameterized by $n$, we impose the constraints $\alpha = 0$ and $\beta_3 \geq 0$ on the weights. By applying all the constraints in Appendix A, we derive the following

inequalities:

$$0 \leq \frac{3\beta_1}{4} \leq \frac{3\beta_2}{4} \leq \beta_3$$

To generate algorithms parameterized by $k$, we impose the constraints $\beta_i \leq 0$ for $i \in \{1, 2\}$ and $\beta_3 = 0$ on the weights. By applying all the constraints in Appendix A, we derive the following inequalities:

$$-\frac{\alpha}{2} \leq \beta_2 \leq -\frac{\alpha}{3}, \qquad\qquad -\frac{\alpha}{2} - \frac{\beta_2}{2} \leq \beta_1 \leq \frac{\alpha}{2} + \frac{3\beta_2}{2}.$$

## 4.3 Branching process: generation, evaluation, and rule formation

In this subsection, we examine how the framework selects the most suitable branching rule for a given local configuration. As detailed in Subsubsection 4.3.1, the process begins by generating candidate branches. These candidates are then evaluated for efficacy and correctness in Subsubsection 4.3.2. Based on this evaluation, Subsubsection 4.3.3 determines an appropriate subset of branches to construct the final branching rule.

Before generating a branching rule, the framework first verifies whether any simplification rules can be applied to the local configuration. If no simplification is possible, it proceeds with the process outlined in this subsection. To clarify how the framework functions, we now introduce several key definitions that underpin this procedure.

▶ **Definition 5** (Branch). *Given a local configuration $L = (H, D)$, a branch $b \subseteq V$ is a set of vertices chosen to be included in the vertex cover. Applying $b$ to $L$ removes these vertices (and their incident edges) from the graph, producing a new local configuration $b(L) = (H', D')$, where $H' = (V', E') = H - b$. The incomplete edge set is updated by keeping $D'(v) = D(v)$ for all $v \in V'$[2].*

▶ **Definition 6** (Branching rule). *A branching rule, denoted by $r$, is a function that maps a local configuration to a set of local configurations. The branching rule $r$ generates this set by applying a set of branch functions to the input local configuration.*

▶ **Definition 7** ($\mu$-efficient branching rule). *For a local configuration $L$ and a measure $\mu$, a branching rule $r$ is said to be $\mu$-efficient if, for all instances in the instance space represented by $L$, it satisfies the constraints imposed by the Measure & Conquer analysis (Lemma 1 for deterministic or Lemma 2 for randomized). Note that we consider simplification rules to be branching rules with exactly one branch.*

### 4.3.1 Branch generation

Let $L = (H, D) \in \texttt{Expand}(L', \Delta)$ be a local configuration obtained from $L' = (H', D')$ by applying the expansion algorithm (Algorithm 2) and $H = (V, E), H' = (V', E')$. We denote the set of candidate branches for $L'$ as $B'$, and our goal is to construct candidate branches for $L$ based on $B'$. For the base case, all subsets of $V'$ are considered as candidate branches.

---

[2] Removing $b$ deletes only edges incident to vertices in $b$. Because the exact endpoints of the deleted incomplete edges are unknown, we approximate by assuming that the number of incomplete edges incident to the remaining vertices does not decrease. This approximation is not strictly safe: in reality some incomplete edges may disappear, so our approximation may overestimates the measure decrease. We introduce a compensation term later to correct this effect (see Lemma 12).

Let $\{u, v\} \in E \setminus E'$ be the unique edge added during the expansion. We then generate a set of candidate branches for $L$, and will later eliminate redundant ones:

$$B = \{b \mid b \in B'\} \cup \{b \cup \{u\} \mid b \in B'\} \cup \{b \cup \{v\} \mid b \in B'\} \cup \{b \cup \{u, v\} \mid b \in B'\}.$$

In essence, this construction reflects the intuition that, since $L$ differs from $L'$ only by the addition of edge $\{u, v\}$, any branch that is suboptimal for $L'$ is likely to remain suboptimal for $L$. Therefore, we systematically account for the new edge by considering the scenarios where $u$, $v$, or both are added to each existing candidate branch.

### 4.3.2 Branch evaluation

Once we have the expanded set of candidate branches for $L$, we next need to prune and evaluate them to find an efficient branching rule. This subsection, we discuss techniques for evaluating the performance of each candidate branch on a local configuration, focusing on two key aspects: correctness and efficiency. To assess correctness, we introduce the concept of a boundary requirement, which determines a set of instances in the instance space that a branch satisfies. For efficiency, we define the branch cost to quantify the progress on measure made by a branch.

▶ **Definition 8** (Boundary requirement). *Consider a local configuration $L = (H, D)$ with $H = (V, E)$. A boundary requirement $R \subseteq \delta(L)$ is a subset of boundary vertices that specifies how the external structure constrains the choice of a minimum vertex cover within the local configuration. More precisely, for any vertex $v \in \delta(L)$, if $v \in R$, then $v$ is required to be included in the vertex cover (so that it covers its incident incomplete edges); if $v \notin R$, then all of $v$'s incomplete edges are assumed to be covered by other vertices.*

A branch $b$ of $L$ *satisfies* the boundary requirement $R$ if $b$ is a subset of a minimum vertex cover after applying the boundary requirement on the local configuration. Formally, $\text{VC}(H - R) = \text{VC}(H - R - b) + |b \setminus R|$.

In order to optimize the branching process, it is essential to reduce the set of boundary requirements under consideration. Since satisfying one boundary requirement can imply the satisfaction of others, we can identify a reduced subset that still guarantees the correctness of the branching process. Consequently, by focusing on a minimal yet sufficient set, we streamline computations without compromising accuracy.

Let $\mathcal{R}_0 = 2^{\delta(L)}$ denote the universe of boundary requirements for a local configuration $L$. We now define the crucial boundary requirement set as follows.

▶ **Definition 9** (Crucial boundary requirement set). *Let $L$ be a local configuration. We define the set of boundary requirements $\mathcal{R}$ as a* crucial boundary requirement set *if every branch that satisfies a boundary requirement $R \in \mathcal{R}_0 \setminus \mathcal{R}$ also satisfies at least one boundary requirement in $\mathcal{R}$.*

It is easy to see that $\mathcal{R}_0$ is a crucial boundary requirement set.

A smaller crucial boundary requirement set can be derived by removing unnecessary elements from $\mathcal{R}_0$. Let $G_{\mathcal{R}_0}$ be a directed acyclic graph where each boundary requirement $R \in \mathcal{R}_0$ corresponds to a node. Consider two boundary requirements $R$ and $R \cup \{v\}$ that differ only by the presence of a single vertex $v$. If including $v$ as required does not increase the minimum vertex cover size (formally, $\text{VC}(H - R) = \text{VC}(H - R - v) + 1$), then we draw a directed edge from node $R \cup \{v\}$ to node $R$ in $G_{\mathcal{R}_0}$. If instead requiring $v$ to be included increases the minimum vertex cover size, we direct an edge from $R$ back to $R \cup \{v\}$. In $G_{\mathcal{R}_0}$, the nodes with no incoming edges form the reduced set $\mathcal{R}_1$.

▶ **Lemma 10.** *For any local configuration, $\mathcal{R}_1$ is a crucial boundary requirement set.*

Algorithm 3 evaluates the correctness of a branch $b$ on a local configuration $L$ by determining which boundary requirements are satisfied by $b$.

**Algorithm 3** $\texttt{EB}(L, b, \mathcal{R})$

---
**Input:** A local configuration $L = (H, D)$, a branch $b$, and a crucial boundary
           requirement set $\mathcal{R}$ for $L$
**Output:** The set of boundary requirement $A$ that $b$ satisfies

1 $A \leftarrow \emptyset$;
2 **foreach** $R_i \in \mathcal{R}$ **do**
3    **if** $\texttt{VC}(H - R_i) = \texttt{VC}(H - R_i - b) + |b \setminus R_i|$ **then**
4       $A \leftarrow A \cup \{R_i\}$;
5 **return** $A$;

---

To apply the M&C method in our framework, we bound the worst-case relative change of the measure when a branch is applied to any instance whose graph contains a given local configuration as an expansion. We capture this change via the following notion of *branch cost*, which quantifies the progress achieved by a branch on that local configuration.

▶ **Definition 11** (Branch cost). *Given a measure $\mu$, the* branch cost *of a branch $b$ on an instance $I$ is the ratio between the measure after and before applying the branch:*

$$\texttt{cost}(I, b) = 2^{\mu(b(I)) - \mu(I)}. \tag{4}$$

*Intuitively, $\texttt{cost}(I, b)$ is the multiplicative shrinkage factor of the measure induced by $b$ on $I$.*

*For a local configuration $L$, the branch cost of $b$ is the maximum over all instances $I = (G, k)$ whose graph expands $L$:*

$$\texttt{cost}(L, b) = \max_{\substack{I = (G, k) \\ G \geq L}} \texttt{cost}(I, b) = \max_{\substack{I = (G, k) \\ G \geq L}} 2^{\mu(b(I)) - \mu(I)}. \tag{5}$$

Let $b$ be a branch for a local configuration $L = (H, D)$ with $H = (V, E)$. Applying $b$ to $L$ yields a new local configuration $L' = b(L) = (H', D')$. We denote by $\Delta k = -|b|$ the change in the vertex cover budget caused by selecting the vertices in $b$, and by $\Delta n_i$ the increase in the number of degree-$i$ vertices from $L$ to $L'$.

Figure 1 illustrates an example of how the measure changes after applying a branch to a local configuration. However, the expression $\alpha \Delta k + \beta_1 \Delta n_1 + \beta_2 \Delta n_2 + \beta_3 \Delta n_3$ can overestimate the actual measure drop if certain structural changes are not accounted for. In particular, when branch $b$ alters the degrees of boundary vertices or affects vertices outside $V$, the measure might not decrease as much as the naive sum suggests. These untracked effects can make the true measure decrease smaller than the above sum, potentially violating the cost estimate's correctness. To address this, we provide a conservative upper bound on the measure change by considering the worst-case scenario in which overestimation is most likely to arise.

Let $\delta_r(L)$ denote the subset of boundary vertices that are not selected into the vertex cover, while $\delta_d(L)$ denotes those that are selected by the branch $b$. We define

$$r_{ij} = \left|\{v \in \delta_r(L) \mid \deg_{H'}(v) + D'(v) = i \wedge D'(v) = j\}\right| \quad \text{and}$$
$$d_{ij} = \left|\{v \in \delta_d(L) \mid \deg_H(v) + D(v) = i \wedge D(v) = j\}\right|.$$

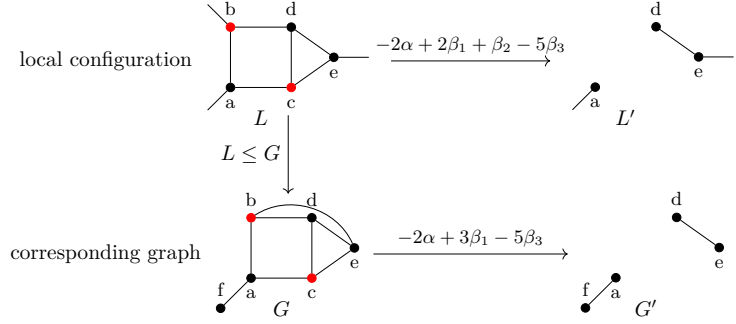

**Figure 1** Measure change on $I = (G, k)$ may be greater than $\alpha \Delta k + \beta_1 \Delta n_1 + \beta_2 \Delta n_2 + \beta_3 \Delta n_3$ of $L$.

▶ **Lemma 12.** *The cost of applying b to the local configuration L is bounded by*

$$\texttt{cost}(L, b) \leq 2^{\alpha\Delta k + \beta_1 \Delta n_1 + \beta_2 \Delta n_2 + \beta_3 \Delta n_3 + \max\left(0,\, (d_{31} + 2d_{32} + \min(d_{21}, r_{21} + 2r_{22} + r_{11}))\max(\beta_1 - \beta_2, -\beta_1)\right)}.$$

However, this estimation can be tightened when certain structural properties of the instance are known. In particular, Lemma 13 and Lemma 14 establish more precise estimations under additional assumptions.

▶ **Lemma 13.** *If there is no instance containing a degree-3 vertex with at least 2 degree-2 neighbors, then the cost of applying b to the local configuration L is bounded by*

$$\texttt{cost}(L, b) \leq 2^{\alpha\Delta k + \beta_1\Delta n_1 + \beta_2\Delta n_2 + \beta_3\Delta n_3 + \max(0, (d_{31} + d_{32} + \min(d_{32} + d_{21}, r_{21} + 2r_{22} + r_{11}))\max(\beta_1 - \beta_2, -\beta_1))}.$$

▶ **Lemma 14.** *If there is no instance containing a degree-2 vertex, then the cost of applying b to the local configuration L is bounded by*

$$\texttt{cost}(L, b) \leq 2^{\alpha\Delta k + \beta_1\Delta n_1 + \beta_2\Delta n_2 + \beta_3\Delta n_3 + \max(0, \min(d_{31} + 2d_{32} + d_{21}, r_{21} + 2r_{22} + r_{11})\max(\beta_1 - \beta_2, -\beta_1))}.$$

We incorporate a post-processing step to filter out branches that are dominated by others, thereby reducing the number of branches considered during the branching rule generation process. Specifically, for any two branches $b$ and $b'$ in the set of candidate branches $B$, if $\texttt{cost}(L, b) \geq \texttt{cost}(L, b')$ and $\texttt{EB}(L, b, R_1) \subseteq \texttt{EB}(L, b', R_1)$, then $b$ is removed from $B$. This pruning process is repeated iteratively until no branch is dominated by another.

Moreover, the reduced set of branches will also serve as the basis for generating candidate branches in future expansions, should no $\mu$-efficient branching rule be identified for the current local configuration.

### 4.3.3 Branching rule generation

Given a set of branches $B$ for a local configuration $L$, a measure $\mu$, and a crucial boundary requirement set $\mathcal{R}$ for $L$, we generate a randomized branching rule using linear programming, as shown in Equation 6. For each branch $b_i \in B$, let $\mathcal{R}_i = \texttt{EB}(L, b_i, R_1)$ represent the set of boundary requirements satisfied by $b_i$. We associate a variable $w_i$ with each branch $b_i$ to represent its weight, as described in Lemma 2. The objective is to minimize the total cost $\sum_i \texttt{cost}(L, b_i) \cdot w_i$, subject to the constraint that every boundary requirement in $\mathcal{R}$ is satisfied by branches whose weights sum to at least one. To generate a deterministic branching algorithm, we restrict each $w_i$ to binary values, i.e., $w_i \in \{0, 1\}$, thereby transforming the problem into an integer linear program.

$$
\begin{aligned}
\underset{\mathbf{w}}{\text{minimize}} \quad & \sum_{b_i \in B} w_i \cdot \text{cost}(L, b_i) \\
\text{subject to} \quad & \forall_{R_i \in \mathcal{R}} \sum_{b_j \in B \text{ s.t. } R_i \in \text{EB}(L, b_j, \mathcal{R})} w_j \geq 1 \\
& \forall_{b_i \in B} w_i \in [0, 1].
\end{aligned}
\quad (6)
$$

▶ **Lemma 15.** *There exists a $\mu$-efficient branching rule $r$ composed of branches in $B$ for $L$ if and only if there exists a solution $w$ to Equation 6 with an objective value $\sum_{b_i \in B} w_i \cdot \text{cost}(L, b_i) \leq 1$.*

## 4.4 Algorithm overview

In this section, we present an overview of the automated framework that leverages this linear programming approach. As detailed in Algorithm 4, the procedure generates a randomized branching algorithm with a running time of $O^*(2^{\mu(I)})$ for a given measure $\mu$. The framework begins with a local configuration $L$, and all graphs in the instance space are expansions of $L$. It first checks for applicable simplification rules; if one is found, it is directly assigned to $L$. Otherwise, the framework generates candidate branches, collects the associated boundary requirements, and then solves the linear program defined in Equation 6 to derive a suitable branching rule. If the resulting rule is $\mu$-efficient, it is adopted for $L$; if not, $L$ is expanded into new local configurations, which are handled recursively. Once this process terminates, the final algorithm is fully specified. For any input instance, the generated algorithm locates the relevant local configuration, applies the corresponding branching rule, and recursively solves one of the resulting sub-instances, selecting it according to the probability distribution given in Lemma 2. This systematic procedure guarantees an efficient branching algorithm tailored to the given measure. Furthermore, to obtain a deterministic branching algorithm, we can enforce binary weights $w_i \in \{0, 1\}$ for all branches in the branching rule.

**Algorithm 4** `GenSA`$(L, \mu, \Delta)$

**Input:** A local configuration $L$, a measure $\mu$, and maximum degree $\Delta$
**Output:** A function $\mathcal{B}$ maps each local configuration to a $\mu$-efficient branching rule

1. **if** *there exists a simplification rule $r$ that can be applied to $L$* **then**
2.      $\mathcal{B}(L) \leftarrow r$;
3.      **return**;
4. $B \leftarrow$ generate branches of $L$;
5. $\mathcal{R} \leftarrow$ collect all boundary requirements of $L$;
6. $w \leftarrow \text{ILP}(L, B, \mathcal{R}, \text{cost}(\cdot), \text{EB}(\cdot))$; // Solve Equation 6
7. **if** $\sum_{b_i \in B} w_i \cdot \text{cost}(L, b_i) \leq 1$ **then**
8.      $r \leftarrow \{w_i b_i \mid b_i \in B\}$;
9.      $\mathcal{B}(L) \leftarrow r$;
10. **else**
11.      **foreach** $L_i \in \text{Expand}(L, \Delta)$ **do**
12.          `GenSA`$(L_i, \mu, \Delta)$;

## 5 Running Time Analysis

In this section, we first present the improved randomized running time upper bounds of VERTEX COVER on subcubic graphs, derived using our generation framework. Subsequently, we leverage the improved bounds in the algorithm by Xiao and Nagamochi [24] to obtain a tighter upper bound on the running time for VERTEX COVER on degree-4 graphs. Finally, we incorporate our bounds into the algorithm of Harris and Narayanaswamy [12], achieving a faster running time for VERTEX COVER on general graphs parameterized by the solution size $k$. Throughout this section, we refer to graphs with maximum degree $i$ as *degree-i graphs*.

To achieve these results, we employ the generation framework described in Section 4 for subcubic graphs, using the measures $\mu_1 = 0.106 n_3$ and $\mu_2 = 0.178k - 0.0445 n_1 - 0.089 n_2$. To enhance efficiency, we partition the instance space of VERTEX COVER on subcubic graphs into 19 subspaces as detailed in Appendix C, and design tailored branching rules for each subspace.[3] The resulting algorithm achieves the following performance:

▶ **Lemma 16.** *There exist randomized algorithms for* VERTEX COVER *on subcubic graphs running in $O^*(1.07625^n)$ time and $O^*(1.13132^k)$ time, respectively.*

As demonstrated by Xiao and Nagamochi [24], their algorithm for INDEPENDENT SET on degree-4 graphs relies on an algorithm for INDEPENDENT SET on degree-3 graphs as a subroutine. Noting that INDEPENDENT SET is complementary to VERTEX COVER, our improved result for VERTEX COVER implies that INDEPENDENT SET on degree-3 graphs can be solved in $O^*(1.07625^n)$ time. By substituting this into Xiao and Nagamochi's framework for degree-4 graphs, we obtain:

▶ **Lemma 17.** *There is a randomized $O^*(1.13735^n)$-time algorithm for* VERTEX COVER *on degree-4 graphs.*

Next, we consider the FPT algorithm for VERTEX COVER proposed by Harris and Narayanaswamy [12]. This algorithm uses a measure of the form $a\mu + bk$, where $\mu$ denotes the gap between the solution size and its LP relaxation. It resolves the case for degree-$i$ graphs relying on two key components: an exact algorithm for VERTEX COVER on degree-$i$ graphs and a recursive application of the same FPT algorithm to degree-$(i-1)$ graphs. This recursive structure applies for degrees $3 \leq i \leq 8$. By incorporating our improved bounds into this framework, we obtain the following result:

▶ **Lemma 18.** *There is a randomized $O^*(1.25281^k)$-time algorithm for* VERTEX COVER *on general graphs. Moreover, for bounded maximum degree: degree-4: $O^*(1.21103^k)$, degree-5: $O^*(1.24382^k)$, degree-6: $O^*(1.25210^k)$.*

To establish Lemma 18, we rely on the following key result:

▶ **Lemma 19** (Lemma 10 of [12])**.** *Suppose that $\mathcal{G}$ is a class of graphs closed under vertex deletion, and for which there are* VERTEX COVER *algorithms with runtimes $O^*(e^{a\mu + bk})$ and $O^*(e^{cn})$ for $a, b, c \geq 0$. Then we can solve* VERTEX COVER *in $\mathcal{G}$ with runtime $O^*(e^{dk})$ where $d = \frac{2c(a+b)}{a+2c}$.*

---

[3] The source code is publicly available at
https://github.com/algo-unsw/A-Faster-Randomized-Algorithm-for-Vertex-Cover. Graph isomorphism detection is supported by Nauty [16], and the linear programming solver is provided by ALGLIB [1].

**Proof of Lemma 18.** We re-examine the `Branch4`–`Branch8` algorithms from Harris and Narayanaswamy [12], integrating our refined running times for VERTEX COVER on degree-3 and degree-4 graphs. Specifically, we obtain:

Algorithm `Branch4` has measure $a_4\mu + b_4k$ with $a_4 = 0.59303$, $b_4 = 0.03958$.
Algorithm `Branch5` has measure $a_5\mu + b_5k$ with $a_5 = 0.37997$, $b_5 = 0.09725$.
Algorithm `Branch6` has measure $a_6\mu + b_6k$ with $a_6 = 0.16828$, $b_6 = 0.16570$.
Algorithm `Branch7` has measure $a_7\mu + b_7k$ with $a_7 = 0.02580$, $b_7 = 0.21576$.
Algorithm `Branch8` has measure $b_8k$ with $b_8 = 0.22539$.

By applying Lemma 19 alongside the running times summarized in Table 1, we derive the stated bounds. ◀

## 6 Conclusion

This work takes a step toward turning hand crafted branching design into a search-and-optimize pipeline. On the analysis side, we extend Measure & Conquer to randomized branching, providing a principled way to upper bound running times when choices are probabilistic rather than deterministic. On the rule generation side, we cast branch discovery and weighting as LP/ILP objectives that directly optimize measure progress, so the system searches a large design space and returns correctness guaranteed, high efficiency rules.

Demonstrated on VERTEX COVER, the pipeline delivers state-of-the-art bounds across several regimes (see Table 1). For subcubic graphs we obtain the fastest known randomized exact and FPT algorithms, with running times $O^*(1.07625^n)$ and $O^*(1.13132^k)$, respectively. As a plug-in improvement, combining our framework with Xiao–Nagamochi [24] yields $O^*(1.13735^n)$ for degree-4 graphs, and pairing with Harris–Narayanaswamy [12] gives $O^*(1.25281^k)$ for general graphs.

Our current implementation assumes a user specified measure and corresponding weights and only checks their feasibility. As measures grow more complex, identifying suitable weights becomes increasingly difficult. A natural next step is to adjust weights automatically (within a prescribed measure family) as configurations expand. Another challenge lies in extending the framework to directly solve the VERTEX COVER problem on graphs with higher degrees, which necessitates handling more configurations and developing smarter strategies for generating branching rules. Finally, the framework could be applied to a broader range of subset problems, ideally those with local structures involving a bounded number of elements that interact with external elements. Such properties help constrain the growth of local configurations and limit the number of boundary requirements, thereby enhancing the scalability of the approach.

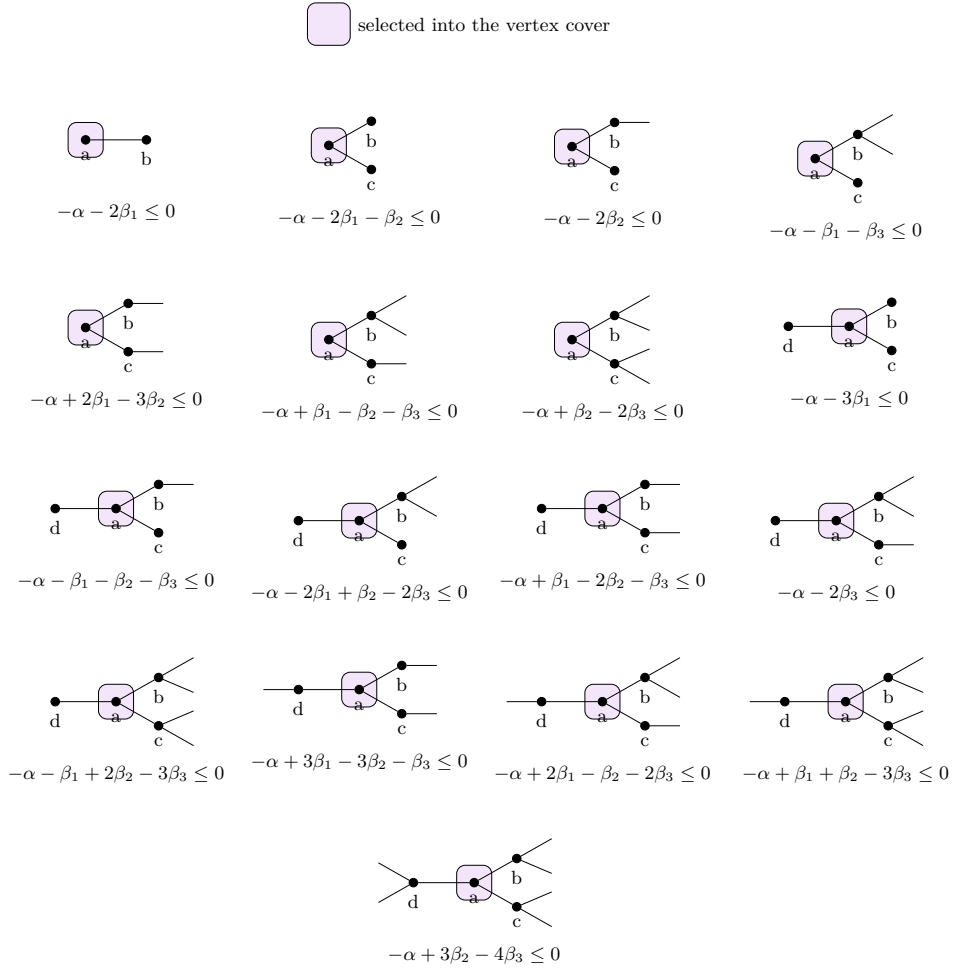

**Figure 2** Atomic operations on subcubic graphs

## A  Measure constraints

Figure 2 and Figure 3 illustrate all operations that can be made on subcubic graphs by simplification rules.

## B  Proofs

▶ **Lemma 2.** *Let $\mathcal{A}_r$ be a randomized branching algorithm for a decision problem $\Pi$, and let $\mu(\cdot)$ be a non-negative measure for instances of $\Pi$. For any instance $I$ of $\Pi$, $\mathcal{A}_r$ either solves $I$ in constant time if $\mu(I) = 0$, or reduces $I$ to sub-instances $I_1, \ldots, I_k$ with corresponding weights $w_1, \ldots, w_k$. If $I$ is a yes-instance, the sum of the weights of the yes-sub-instances is at least one; otherwise, all sub-instances are no-instances. A sub-instance $I_i$ is selected at random with probability at least $w_i \cdot 2^{\mu(I_i) - \mu(I)}$, subject to the constraint*

$$\sum_{i=1}^{k} w_i \cdot 2^{\mu(I_i)} \leq 2^{\mu(I)}. \tag{2}$$

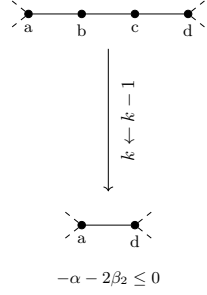

**Figure 3** Applying the simplification rule 4

*If $\mathcal{A}_r$ recursively solves the chosen sub-instance $I_i$ in polynomial depth and uses its solution to solve $I$, with polynomial time required for the reduction and backtracking, then $\mathcal{A}_r$ runs in polynomial time on every instance. Moreover, for any yes-instance $I$, $\Pr[\mathcal{A}_r \text{ solves } I] \geq 2^{-\mu(I)}$.*

**Proof.** We begin by observing that, for any instance $I$, the total probability lower bound for selecting a sub-instance is bounded above by 1:

$$\sum_{i=1}^{k} w_i 2^{\mu(I_i) - \mu(I)} = \sum_{i=1}^{k} \frac{w_i 2^{\mu(I_i)}}{2^{\mu(I)}} \leq 1.$$

Next, we establish the algorithm's success probability by induction on $\mu(I)$. For the base case, note that when $I$ is a constant-size yes-instance, the success probability satisfies $1 \geq 2^{-\mu(I)}$ trivially. Now, assume that for any instance $I'$ with $0 \leq \mu(I') \leq \mu(I) - 1$ and $I'$ being a yes-instance, the inductive hypothesis holds.

Then, for any yes-instance $I$, let $s_i = 1$ if $I_i$ is a yes-instance and 0 otherwise, the success probability satisfies:

$$P_{\mathcal{A}_r}(I) \geq \sum_{i=1}^{k} w_i 2^{\mu(I_i) - \mu(I)} s_i P_{\mathcal{A}_r}(I_i) \qquad \text{(assuming } I_i \text{ is a yes-instance)}$$

$$\geq \sum_{i=1}^{k} w_i 2^{\mu(I_i) - \mu(I)} s_i 2^{-\mu(I_i)} \qquad \text{(by the inductive hypothesis)}$$

$$= 2^{-\mu(I)} \sum_{i=1}^{k} w_i s_i$$

$$\geq 2^{-\mu(I)}. \qquad \text{(the sum of } w_i \text{ of yes-sub-instances is at least 1)}$$

Therefore, the inductive step holds, ensuring that the success probability lower bound is maintained for all yes-instances. ◀

▶ **Lemma 10.** *For any local configuration, $\mathcal{R}_1$ is a crucial boundary requirement set.*

**Proof.** Since $\mathcal{R}_0$ is a crucial boundary requirement set, suppose there exists a branch $b$ that satisfies some boundary requirement $R \in \mathcal{R}_0 \setminus \mathcal{R}_1$. We aim to show that $b$ also satisfies a boundary requirement $R' \in \mathcal{R}_1$.

By Proposition 20, if a branch satisfies a boundary requirement, it necessarily satisfies all boundary requirements that are successors of that requirement in the directed graph $G_{\mathcal{R}_0}$.

Moreover, since $G_{\mathcal{R}_0}$ is a directed acyclic graph, every boundary requirement in $\mathcal{R}_0 \setminus \mathcal{R}_1$ must have at least one predecessor in $\mathcal{R}_1$. Therefore, $b$ must satisfy some predecessor $R' \in \mathcal{R}_1$. ◂

▶ **Lemma 12.** *The cost of applying b to the local configuration L is bounded by*

$$\texttt{cost}(L,b) \leq 2^{\alpha \Delta k + \beta_1 \Delta n_1 + \beta_2 \Delta n_2 + \beta_3 \Delta n_3 + \max\left(0,\, (d_{31} + 2d_{32} + \min(d_{21}, r_{21} + 2r_{22} + r_{11}))\max(\beta_1 - \beta_2, -\beta_1)\right)}.$$

**Proof.** For any instance $I = (G, k)$ with graph $G \geq L$, applying $b$ to $I$ produces a new graph $G' \geq L'$ and decrease $k$ by $|\Delta k|$. The changes in $n_1$, $n_2$, and $n_3$ can be analyzed by considering both the visible changes ($\Delta n_1$, $\Delta n_2$, $\Delta n_3$) and the invisible changes that occur within $I$ but are not explicitly reflected in $L$. These invisible changes fall into two main categories: degree changes for vertices in $\delta(L)$ and degree changes for vertices outside $V$.

In the case of $\delta(L)$, incomplete edges connected to boundary vertices that appear in $G$ may be eliminated when placing a boundary vertex into the vertex cover. This removal can alter the degrees of vertices incident to these edges, potentially resulting in fewer remaining degree-1 ($\Delta n_1$) or degree-2 ($\Delta n_2$) vertices than initially expected. Consequently, the measure decrease by applying $b$ to $I$ may be less than $|\alpha \Delta k + \beta_1 \Delta n_1 + \beta_2 \Delta n_2 + \beta_3 \Delta n_3|$.

Since a vertex decreasing from degree-3 to degree-2 (or lower) does not increase the measure, it is sufficient to focus on cases where an incomplete edge connected to a degree-1 or degree-2 vertex is removed. In total, there are $d_{31} + 2d_{32} + d_{21}$ such incomplete edges: $d_{31}$ edges connected to a degree-3 vertex with one incomplete edge in $\delta_d(L)$, $2d_{32}$ edges connected to degree-3 vertices with two incomplete edges in $\delta_d(L)$, and $d_{21}$ edges connected to degree-2 vertices with one incomplete edge in $\delta_d(L)$.

For an incomplete edge attached to a degree-3 vertex, the other endpoint may be a degree-2 or degree-1 vertex in $\delta_r(L)$ or outside the local configuration, increasing the measure by at most $\max(\beta_1 - \beta_2, -\beta_1)$. Therefore, there are at most $d_{31} + 2d_{32}$ degree changes, each potentially reducing a degree-2 vertex to degree-1 or a degree-1 vertex to degree-0, contributing up to $(d_{31} + 2d_{32}) \cdot \max(\beta_1 - \beta_2, -\beta_1)$ to the measure.

Similarly, for incomplete edges connected to degree-2 vertices, each edge may attach to a degree-2 or degree-1 vertex in $\delta_r(L)$, increasing the measure by at most $\max(\beta_1 - \beta_2, -\beta_1)$. Notably, these edges cannot connect to a degree-2 vertex outside $V$ since such structures are handled by Simplification Rule 4. Here, at most $\min(d_{21}, r_{21} + 2r_{22} + r_{11})$ degree changes can occur, where $r_{21}$ vertices may drop from degree-2 to degree-1, $r_{22}$ from degree-2 to degree-0 via two removed edges, and $r_{11}$ from degree-1 to degree-0. Each change contributes up to $\max(\beta_1 - \beta_2, -\beta_1)$, yielding a total increase bounded by $\min(d_{21}, r_{21} + 2r_{22} + r_{11}) \cdot \max(\beta_1 - \beta_2, -\beta_1)$.

Therefore, these bounds collectively ensure that the lower bound on the measure decrease resulting from applying $b$ to $L$ holds as required. ◂

▶ **Lemma 13.** *If there is no instance containing a degree-3 vertex with at least 2 degree-2 neighbors, then the cost of applying b to the local configuration L is bounded by*

$$\texttt{cost}(L,b) \leq 2^{\alpha \Delta k + \beta_1 \Delta n_1 + \beta_2 \Delta n_2 + \beta_3 \Delta n_3 + \max(0, (d_{31} + d_{32} + \min(d_{32} + d_{21}, r_{21} + 2r_{22} + r_{11}))\max(\beta_1 - \beta_2, -\beta_1))}.$$

**Proof.** Since no instance contains a degree-3 vertex with at least two degree-2 neighbors, a degree-3 vertex can have at most one degree-2 neighbor. Therefore, for any degree-3 vertex with two incomplete edges, at most one of these edges can connect to a degree-2 vertex outside the local configuration in the instance. The other incomplete edge may connect within $\delta_r(L)$. Removing this edge can cause a vertex in $\delta_r(L)$ to decrease its degree by one and a vertex outside the local configuration may decrease its degree by one.

Consequently, removing such incomplete edges may result in an increase in the measure of at most

$$\bigl(d_{31} + d_{32} + \min(d_{32} + d_{21},\, r_{21} + 2r_{22} + r_{11})\bigr) \cdot \max(\beta_1 - \beta_2,\, -\beta_1).$$

◀

▶ **Lemma 14.** *If there is no instance containing a degree-2 vertex, then the cost of applying $b$ to the local configuration $L$ is bounded by*

$$\texttt{cost}(L, b) \leq 2^{\alpha \Delta k + \beta_1 \Delta n_1 + \beta_2 \Delta n_2 + \beta_3 \Delta n_3 + \max(0,\, \min(d_{31} + 2d_{32} + d_{21},\, r_{21} + 2r_{22} + r_{11}) \max(\beta_1 - \beta_2,\, -\beta_1))}.$$

**Proof.** Since no instance contains a degree-2 vertex, all vertices outside $L$ must be degree-3 vertices. Therefore, removing an incomplete edge connected to a vertex outside the local configuration will not increase the measure, as the degree reduction does not affect the measure for degree-3 vertices. However, each removed incomplete edge may still be connected to a vertex in $\delta_r(L)$, and removing such an edge can cause that vertex to decrease its degree by one.

As a result, removing the $d_{31} + 2d_{32} + d_{21}$ incomplete edges may lead to an increase in the measure of at most

$$\min\bigl(d_{31} + 2d_{32} + d_{21},\, r_{21} + 2r_{22} + r_{11}\bigr) \cdot \max(\beta_1 - \beta_2,\, -\beta_1).$$

◀

▶ **Lemma 15.** *There exists a $\mu$-efficient branching rule $r$ composed of branches in $B$ for $L$ if and only if there exists a solution $w$ to Equation 6 with an objective value $\sum_{b_i \in B} w_i \cdot \texttt{cost}(L, b_i) \leq 1$.*

**Proof.** We begin by demonstrating the forward direction. Assume that there exists a solution $w$ to Equation 6 with an objective value satisfying $\sum_{b_i \in B} w_i \cdot \texttt{cost}(L, b_i) \leq 1$. The branching rule $r$ can then be constructed by assigning weight $w_i$ to each branch $b_i$.

To verify the correctness of this branching rule, consider any instance $I = (G, k)$ with boundary requirement $R$ where $G \geq L$. Since Equation 6 holds, the sum of the weights of the sub-instances that satisfy $R$ is at least one.

Next, we establish the efficiency of the branching rule. By definition, each branch $b_i \in r$ incurs a cost $\texttt{cost}(L, b_i)$, and the total weighted cost does not exceed one. Therefore, for any instance $I = (G, k)$ with $G \geq L$, we have:

$$\begin{aligned}
2^{\mu(I)} &\geq 2^{\mu(I)} \sum_{b_i \in r} w_i \cdot \texttt{cost}(L, b_i) && \text{(since the objective value is } \leq 1\text{)} \\
&\geq 2^{\mu(I)} \sum_{b_i \in r} w_i 2^{\mu(I_i) - \mu(I)} && \text{(by the definition of branch cost)} \\
&= \sum_{b_i \in r} w_i 2^{\mu(I_i)},
\end{aligned}$$

which shows that $r$ satisfies the constraints in Lemma 2.

For the backward direction, suppose there exists a $\mu$-efficient branching rule $r$ for $L$ composed of branches from $B$. One can then use the weights $w_i$ of each branch $b_i$ as the variables in Equation 6. Since $\sum_{b_i \in B} w_i \cdot \texttt{cost}(L, b_i) \leq 1$, it follows that the solution to Equation 6 will have an objective value no greater than one.

Therefore, the lemma holds. ◀

▶ **Proposition 20.** *Let $L = (H, D)$ be a local configuration with $H = (V, E)$. For any set of boundary vertices $C \subseteq \delta(L)$ and vertex $v \in \delta(L) \setminus C$, if $\text{VC}(H - C) + |C| = \text{VC}(H - C - v) + |C \cup \{v\}|$, then any branch $b$ that satisfies boundary requirement $C \cup \{v\}$ implies $b$ satisfies boundary requirement $C$. Otherwise, any branch $b$ that satisfies $C$ will also satisfy $C \cup \{v\}$.*

**Proof.** We begin by discussing a fundamental property of the minimum vertex cover. Let $H = (V, E)$ be a graph, for any vertex $v \in V$, the size of the minimum vertex cover satisfies the inequality $\text{VC}(H - v) \leq \text{VC}(H) \leq \text{VC}(H - v) + 1$, since a vertex $v$ in $H$ must either be included in or excluded from a minimum vertex cover of $H$.

To prove the proposition, we need to consider both cases.

First, suppose that $\text{VC}(H - C) + |C| = \text{VC}(H - C - v) + |C \cup \{v\}|$, and a branch $b$ satisfies the boundary requirement $C \cup \{v\}$. Assume $b$ selects $U \subseteq V$ into the vertex cover. We then have:

$$\begin{aligned}
&\text{VC}(H - C - v) + |C \cup \{v\}| \\
=&\text{VC}(H - C - v - U) + |C \cup \{v\} \cup U| &&(b \text{ satisfies boundary requirement } C \cup \{v\}) \\
\geq&\text{VC}(H - C - U) + |C \cup U| &&(\text{by the properties of vertex cover}) \\
\geq&\text{VC}(H - C) + |C|. &&(\text{by the properties of vertex cover})
\end{aligned}$$

Since $\text{VC}(H-C)+|C| = \text{VC}(H-C-v)+|C\cup\{v\}|$, we conclude that $\text{VC}(H-C-U)+|C\cup U| = \text{VC}(H-C)+|C|$. Thus, $b$ satisfies the boundary requirement $C$.

Now, consider the case where $\text{VC}(H - C) + |C| \neq \text{VC}(H - C - v) + |C \cup \{v\}|$. In this case, we have $\text{VC}(H - C) + |C| = \text{VC}(H - C - v) + |C \cup \{v\}| - 1$. Assume that $b$ satisfies the boundary requirement $C$. Then, we have:

$$\begin{aligned}
&\text{VC}(H - C) + |C| \\
=&\text{VC}(H - C - U) + |C \cup U| &&(b \text{ satisfies boundary requirement } C) \\
\geq&\text{VC}(H - C - v - U) + |C \cup \{v\} \cup U| - 1 &&(\text{by the properties of vertex cover}) \\
\geq&\text{VC}(H - C - v) + |C \cup \{v\}| - 1. &&(\text{by the properties of vertex cover})
\end{aligned}$$

Since $\text{VC}(H - C) + |C| = \text{VC}(H - C - v) + |C \cup \{v\}| - 1$, we conclude that $\text{VC}(H - C - v - U) + |C \cup \{v\} \cup U| = \text{VC}(H - C - v) + |C \cup \{v\}|$. Therefore, $b$ satisfies the boundary requirement $C \cup \{v\}$. ◀

## C  Instance Space Partition

This appendix provides the detailed case analysis for subcubic graphs referenced in Section 5.

The instance space of the VERTEX COVER problem on subcubic graphs is systematically partitioned through a structured case analysis. Each subspace is defined by the presence of a specific local structure, while excluding configurations already accounted for in earlier subspaces.

We divide the instance space $\mathcal{I}$ into 19 disjoint subspaces $P_1, \ldots, P_{19}$, based on the presence or absence of specific structures. Each instance falls into exactly one of these subspaces, characterized as follows:

$P_1$: Contains a degree 0 or 1 vertex.

$P_2$: Contains a degree 3 vertex adjacent to at least two degree 2 vertices, excluding instances from $P_1$.

$P_3$: Contains a 4-cycle with three degree 3 vertices and one degree 2 vertex, excluding instances from previous subspaces.

$P_4$: Contains a 5-cycle with four degree 3 vertices and one degree 2 vertex, excluding instances from previous subspaces.

$P_5$: Contains a 6-cycle with five degree 3 vertices and one degree 2 vertex, excluding previous subspaces.

$P_6$: Contains a degree 2 vertex, excluding instances from previous subspaces.

$P_7$: Contains a 3-cycle, excluding previous subspaces.

$P_8$: Contains a 4-cycle, excluding previous subspaces.

$P_9$: Contains two 5-cycles sharing an edge, excluding instances from previous subspaces.

$P_{10}$: Contains a 5-cycle and a 7-cycle sharing an edge, excluding previous subspaces.

$P_{11}$: Contains a 5-cycle, excluding previous subspaces.

$P_{12}$: Contains two 6-cycles sharing an edge, excluding previous subspaces.

$P_{13}$: Contains a 6-cycle, excluding previous subspaces.

$P_{14}$: Contains two 7-cycles sharing three edges, excluding previous subspaces.

$P_{15}$: Contains two 7-cycles sharing two edges, excluding previous subspaces.

$P_{16}$: Contains two 7-cycles sharing one edge, excluding previous subspaces.

$P_{17}$: Contains a 7-cycle, excluding all previous subspaces.

$P_{18}$: Contains an 8-cycle, excluding all previous subspaces.

$P_{19}$: Contains no structures defined by the previous subspaces.

Since these subspaces cover the entire instance space of VERTEX COVER on subcubic graphs, every instance $I \in \mathcal{I}$ belongs to one of them. Given a measure $\mu$, if for every instance $I$ in each subspace $P_i$ there exists a branching rule that is $\mu$-efficient, then we can construct an algorithm that solves any instance $I \in \mathcal{I}$ in $O^*(2^{\mu(I)})$ time.